\documentclass[%
 preprint,
 amsmath,amssymb,
prl,
]{revtex4}

\usepackage{graphicx}
\usepackage{dcolumn}
\usepackage{bm}

\usepackage{xcolor}
\usepackage{soul}
\usepackage{ulem}

\begin{document}


\title{The Abrikosov vortex structure revealed through near-field radiative heat exchange}

\author{S. G. Castillo-L\'opez, R. Esquivel-Sirvent, G. Pirruccio, and C. Villarreal}

\affiliation{%
	Instituto de F\'isica, Universidad Nacional Autónoma de México, Cd. de México C.P. 04510 Mexico.
}%




\date{\today}

\begin{abstract}
One of the signatures of superconductivity is the formation of the Abrikosov vortex lattice in type-II superconductors in the presence of an external magnetic field. Here, we study the near-field radiative heat transfer between a spherical nanoparticle and a nearby planar substrate, both made of optimally-doped YBa$_2$Cu$_3$O$_{7-\delta}$. We show that the heat flux displays a periodic spatial pattern congruent with the material optical response modulated by the Abrikosov lattice. Our results enrich the toolbox of methods available to study non-conventional superconductivity.	
	
\end{abstract}

\maketitle
 
Understanding the fundamental mechanisms governing thermal emission is of primary importance both from a fundamental point of view and practical applications such as thermal management \cite{Transistor}, thermal rectification \cite{yuan2022near,Yue,wen2019ultrahigh,diodes} and energy conversion devices \cite{lucchesi2021near}. At submicron distances, for body separations well below the Wien thermal wavelength, $\lambda_T=\hbar c/k_B T$, the near-field radiative heat transfer (NFRHT) is dominated by the electromagnetic evanescent modes supported by each material \cite{HARGREAVES,vanhove}; therefore, thermal radiation exceeding the black-body limit between contactless materials may constitute the main heat flux channel. In past years, many systems have been considered based on geometric or physical properties \cite{PhysRevB.95.125404,PhysRevB.92.125418}, giving rise to fundamental questions about the transport properties \cite{Francoeur15}, sample size effects \cite{Gelais,gelaisnearfield}, and coherence effects occurring at the micro and nanoscale \cite{jones2013349,yu2023manipulating}.

Superconducting (SC) systems yield the possibility of assessing NFRHT modifications associated to the shift  from normal to quantum coherent behavior for temperatures above and below the critical value $T_c$.
Recently, NFRHT suppression between parallel plates of BCS superconductors has been experimentally shown. 
Similarly, a pronounced suppression or enhancement of NFRHT has been predicted for non-conventional superconductors, particularly, setups involving YBa$_2$Cu$_3$O$_{7-\delta}$ (YBCO), depending on whether  bulk or film configurations are considered,  as well as on changes of the optical response due to the superconducting transition  \cite{castillo2020near,castillo2022enhancing}.  HTSCs offer the opportunity of achieving superfluid charge carrier motion at temperatures considerably higher than conventional BCS superconductors, opening the possibility of their use in heat flux modulation  for applications involving thermal switches or thermal emitters. 

\begin{figure}[t]
	\includegraphics[width=0.45\textwidth]{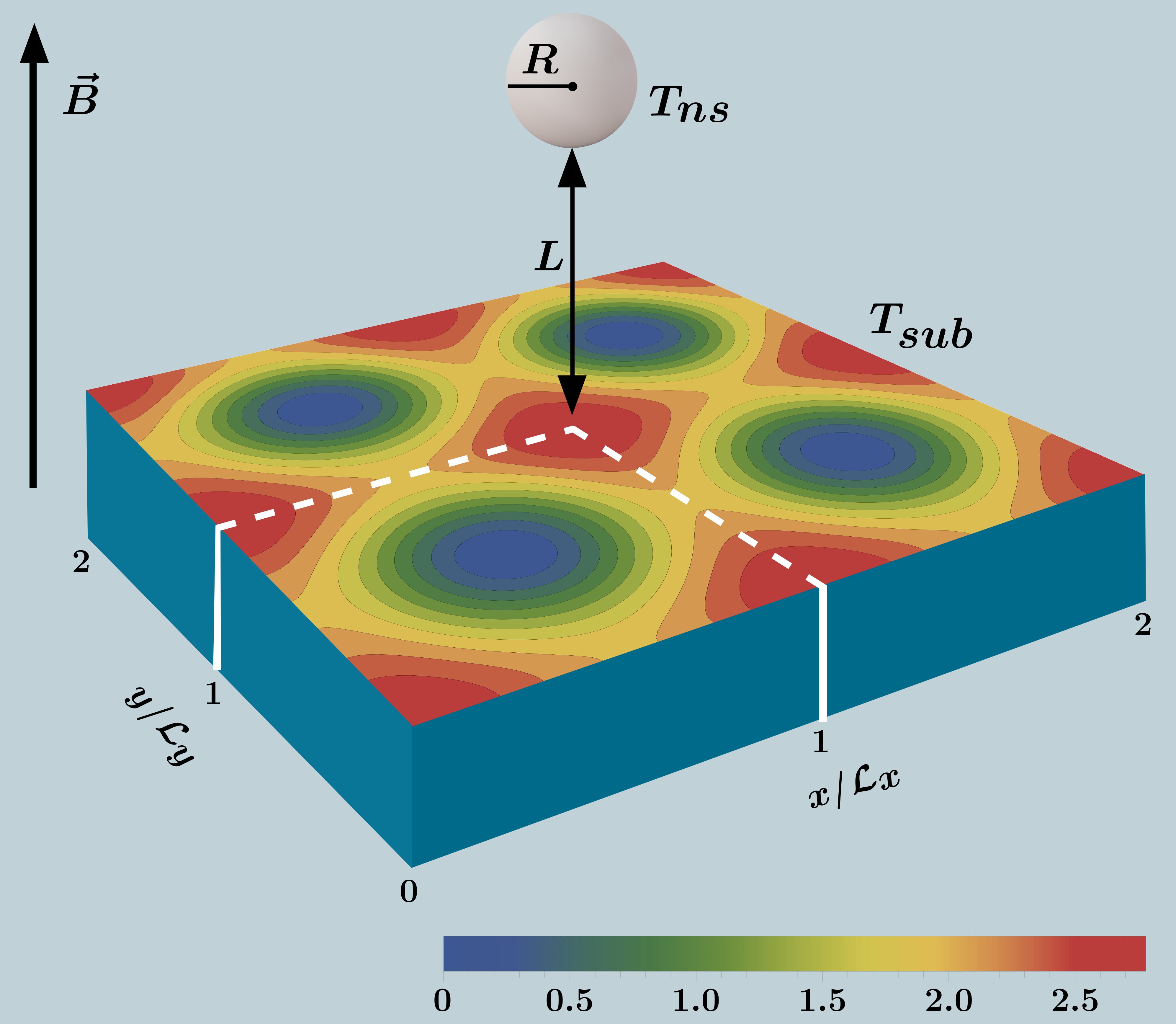}
	\caption{Setup consisting of  a spherical YBCO nanoparticle of radius $R$ in the normal state  ($T_{sph}>T_c$) and a planar YBCO substrate in the SC state ($T_{sub}<T_c$), a distance $L$ apart, in  presence of an external magnetic field. The substrate subtends an  order parameter $n_s^{2D}(\mathbf{r})=\vert\Psi_0(\tilde{x},\tilde{y})\vert^2$ defining a periodic lattice with periods $\mathcal{L}_x=\mathcal{L}_y$. Superconducting current lines  coincide with equiprobability lines of the order parameter. Vortex cores correspond to the darkest zones.}
	\label{fig-01}
\end{figure}

Under the action of the applied field, conventional BCS superconductors exhibit a perfect diamagnetic response, --the Meissner effect-, such that for field intensities, $B<B_c$,  surface supercurrents  expel the magnetic field lines from their interior within a penetration length $\lambda(T)=(4\pi e^{*2} n_s(T)/m^*)^{-1/2}$, where $e^*=2e$, $m^*$, and $n_s(T)$  represent the charge, mass and number  density of Cooper pairs (CPs) at a temperature $T$. In contrast, 
HTSCs involve a mixed diamagnetic phase, delimited by two critical field intensities, $B_{c1}< B_{c2}$, in which normal charge carriers with density $n_n(T)$ coexist with  SC  pairs, such that the charge carrier density can be split as  $n=n_n(T)+n_s(T)$,  In this phase,
\textit{local} screening supercurrents, $\mathbf{J}_s$, force the magnetic flux to penetrate the sample in the form of quantum vortexes  with an intensity  $\Phi_0=h/e^*$ \cite{annett2004superconductivity}. Even though a
field increase beyond $B_{c2}$ causes the disappearance of superconductivity,  Abrikosov found that for intensities of the field just below $B_{c2}$, the quantum vortexes align  displaying a lattice  with periods $\mathcal{L}_x=\mathcal{L}_y=\mathcal{L}$ (see Fig.~\ref{fig-01}) \cite{abrikosov2004nobel}.

In this Letter, we tackle the fundamental question on whether the Abrikosov lattice influences the local thermal properties of HTSCs. Specifically, we study the NFRHT between a spherical YBCO nanoparticle of radius $R$ in the normal state  ($T>T_c$) and a planar YBCO substrate in the SC state ($T<T_c$), in the presence of an external magnetic field. This setup is sketched in Fig.~\ref{fig-01}. By assuming that $R\lesssim \mathcal{L}$, this setup allows  a local scanning of the heat flux. We show that, within a dipolar approximation for the sphere-substrate coupling,  the heat flux acquires a spatial structure inherited from the Abrikosov lattice. Surprisingly, we find that, while for temperatures $T\lesssim T_c$ the heat flux maxima coincide with the vortex core centers,  for temperatures $T\ll T_c$ they  radially migrate, forming a ringed structure over the vortex core borders.   

To characterize the dielectric permittivity of cuprate ceramics, like YBCO, we put forth a variant of the  Ginzburg-Landau (GL) superconductivity theory that considers specific attributes of these materials. In the GL theory, the transit to the SC state is described as a second-order phase transition determined by  a complex order parameter, $\Psi(\mathbf{r})$, null in the normal phase, but finite in the SC phase, implying a long-range order specified by the number density of SC pairs: $n_s(\mathbf{r})=\vert\Psi(\mathbf{r})\vert^2$ \cite{tinkham2004introduction,annett2004superconductivity}.
In presence  of an applied magnetic field, $\mathbf{B}=\nabla \times \mathbf{A}$, the order parameter satisfies the non-linear GL equation
\begin{equation}\label{GL0}
    \left[\frac{1}{2m^*}\left( -i\hbar \nabla-e^*\mathbf{A}/c\right)^2 -(a_T+b \vert \Psi_0\vert^2)\right]\Psi_0=0,
\end{equation}
where $a_T= a_0(T-T_c)$, and $b>0$ in the standard GL approach. In absence of external fields or boundaries, this equation predicts a homogeneous SC ordering, $\vert\Psi_\infty\vert^2=-a_T/b$.  Also, for an interface between a normal metal ($z<0$) and a superconductor ($z>0$), Eq.(\ref{GL0}) yields $\Psi(z)=\Psi_\infty \tanh[z/\sqrt{2} \, \xi(T)]$, where the GL coherence length $\xi(T)=(\hbar^2/2m^* \vert a_T\vert)^{1/2}$ is the distance in which the order parameter acquires its bulk value, $\Psi_\infty$ \cite{tinkham2004introduction}. Given that $\xi(T)$ is microscopically determined by the CP correlation length,
 the parameter $\kappa=\lambda(T)/\xi(T)$ represents a ratio of CP separation with respect to their intrinsic size.

We incorporate into the GL theory two singular features of optimally doped HTSCs:
(\textit{a}) An extremely short pair coherence length, such that  $\kappa \ll1$,  
 congruent with a \textit{local} description of the electromagnetic properties of the system \cite{abrikosov1957magnetic}; moreover, 
 this is indicative of a strongly attractive inter-fermion  interaction, leading to the formation of single-pair bound states able to undergo Bose-Einstein condensation (BEC) \cite{nozieres1985bose}.
(\textit{b}) A  layered  crystallographic structure, in which superfluid charge transport  occurs mainly along  CuO$_2$ planes -the $ab$-planes-. Together, these  conditions imply that YBCO pairs define a 2D dilute boson gas,  with an energy spectrum $E_k=-V_0+\mathcal{E}_k$, where $V_0$ is the pair binding energy, and $\mathcal{E}_k$ the excitation spectrum; in the dilute regime, this may be approximated by a phonon dispersion relation,  $\mathcal{E}_k \approx \hbar c_s k$, with $c_s$ the  sound's speed. BEC formation occurs at a temperature $T_c$ when the pair chemical potential  $2\mu=-V_0$, and the pair occupancy number can be expressed, for $T<T_c$,  as follows:
 \begin{equation}\label{GP}
     n^{2D}(\mathbf{r},T)=n_0^{2D}(\mathbf{r},T) +\sum_{\mathbf{k}>0} \frac{1 }{\exp\left( \hbar c_s k/k_BT\right)-1}
  \end{equation}
 Since the linear structure of  $\mathcal{E}_k$ obeys Landau's criterion for superfluidity \cite{landau2013statistical}, it turns out that  the  density of SC carriers  coincides in this limit with the condensate: $n_s^{2D}(\mathbf{r},T)\approx n_0^{2D}(\mathbf{r},T)$.
Straightforward evaluation of the summation in Eq.(\ref{GP}) then leads to:
\begin{equation}\label{n2d}
      n_s^{2D}(\mathbf{r},T)=\left(1-T^2/T_c^2\right) N \vert\Psi_0(\mathbf{r})\vert^2.
\end{equation}
Here, $ T_c= \left( 2\pi \hbar^2 c_s^2 n^{2D}/k_B^2\zeta(2)\right)^{1/2}$ \cite{fujita1996quantum}, with $\zeta(2)$  a Riemann's zeta function,   $\int d^2 r \vert\Psi_0(\mathbf{r})\vert^2=1$, and $N$  is the number of charge carriers per unit area. The general relation (\ref{n2d}) describes, in absence of magnetic fields,  experimental measurements performed in a number HTSC materials under a diversity of doping levels \cite{zuev2005correlation,chen2005bcs,lomnitz2013bec}. On the other hand,
consistency with the original GL theory is achieved by assuming that the parameter $a_T=-\widetilde{a_0}(1-T^2/T_c^2)$ in Eq.(\ref{GL0}). This implies  in turn that the $ab$-correlation length $\xi_{ab}(T)=\xi_{ab}^2(0)/(1-T^2/T_c^2)$.  

 In the following, we combine relation (\ref{n2d}) with
spectroscopic measurements of the YBCO optical response obtained at the normal ($T=100~\mathrm{K}$), and SC phases ($T=2~\mathrm{K}$) \cite{Bonn1988PhysRevB.37.1574}. As a result, we  obtain an expression for  the $ab$-plane dielectric permittivity in the SC phase, valid  in the whole temperature range $0<T<T_c$:  
\begin{eqnarray}\label{eps}
	\varepsilon^{(s)}_{ab}&=&\varepsilon_\infty-\frac{\omega_{p}^2}{\omega^2}  \left(1-\frac{T^2}{T_c^2}\right)  \vert \Psi_0(\mathbf{r})\vert^2  \left[1+ 
	\frac{i\pi\omega \delta(\omega)}{2} \right] \\ \nonumber
	&-& \left(\frac{T^2}{T_c^2}\right)\frac{\omega_{p}^2}{\omega^2 + i \gamma(T) \, \omega}-\sum_{r=1}^{N_{ph}+1} \frac{\omega_{pr}^2}{\omega^2 - \omega_r^2+  i\gamma_r  \omega}.
\end{eqnarray}
Here, the temperature-dependent terms represent Drude-like contributions, with plasma frequency, $\omega_p$, and scattering rate, $\gamma(T)$, while the summation includes contributions associated to a mid-infrared and $N_{ph}$ phonon resonances.  The normal state permittivity, $\varepsilon_{ab}^{(n)}$, may be derived from the limit $T \to T_c$ of (\ref{eps});  the respective  sphere and substrate permittivities are  then given by $\varepsilon_{ns}=\varepsilon^{(n)}_{ab}$ and $\varepsilon_{sub}=\varepsilon^{(s)}_{ab}$ (see the Supplementary Information (SI) for more details on the YBCO permittivity in the normal and SC phases). To discuss our results, it is worth mentioning that the main contributions to the NFRHT at subwavelength separations, $L<\lambda_T$, arise form evanescent surface modes (SMs) that  may be excited for frequencies such that $\varepsilon_{ab}^{(s)}(\omega)=-1$ \cite{ibach2009solid}.

\begin{figure}[t]
	\includegraphics[width=0.45\textwidth]{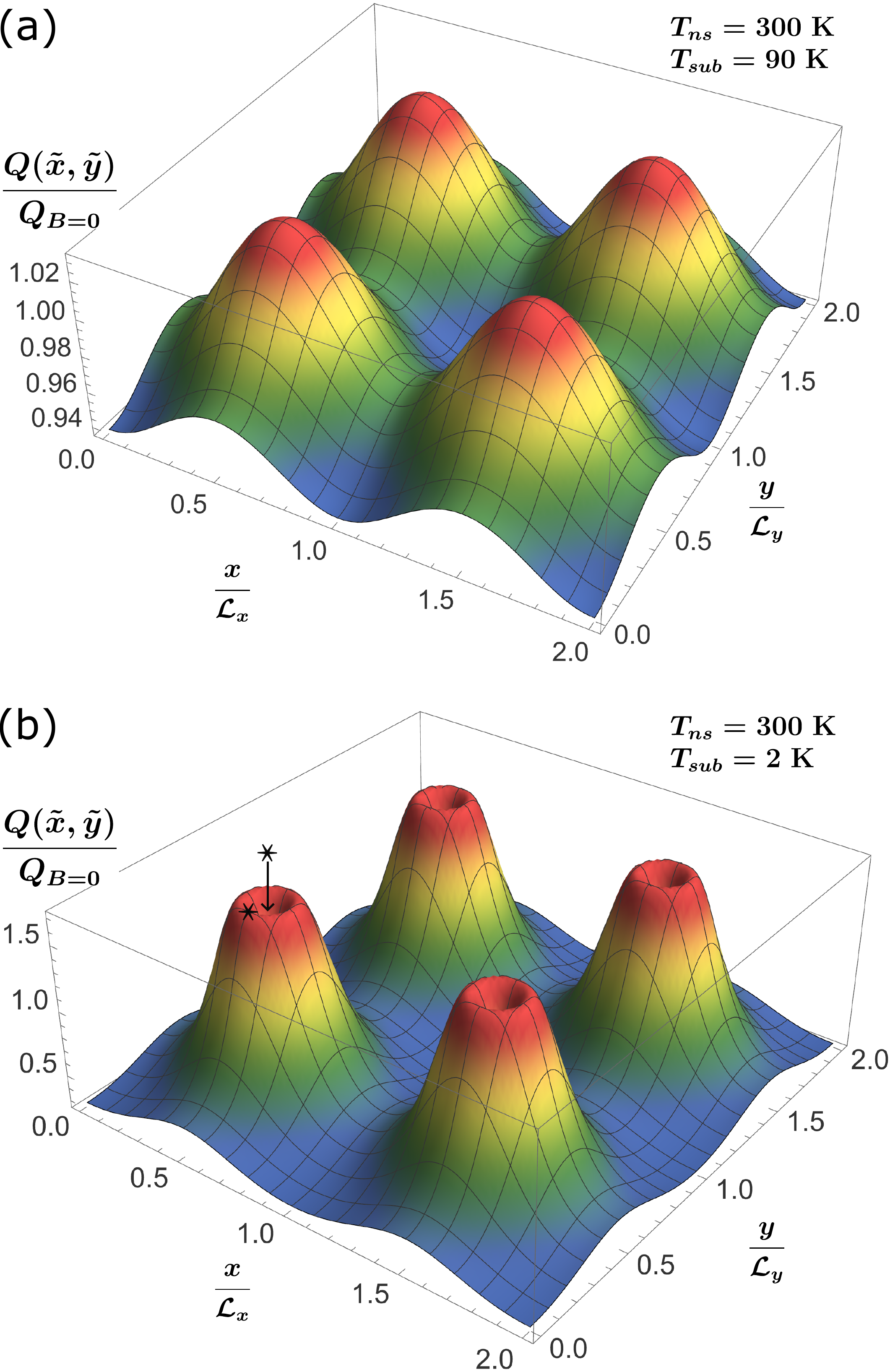}
	\caption{Spatial distribution of heat flux between a nanosphere and a YBCO substrate with an Abrikosov vortex structure. Here $R=10$ nm, $L=20$ nm and, (a) $T_{ns}=300$ K, $T_{sub}=90$ K, (b) $T_{ns}=300$ K, $T_{sub}=2$ K. Accordingly with our theoretical predictions,  $\mathcal{L}_x(2 \, \mathrm{K}) \approx 6$ nm, whereas $\mathcal{L}_x(90 \, \mathrm{K}) \approx 100$ nm. Notice that the vortex size can be tuned using the lattice parameter, $\mathcal{L}_{x,y}$, and the temperature $T_{sub}$. See Supplementary Fig.~S1.}
	\label{fig-02}
\end{figure}

The explicit expression of $\Psi_0(\mathbf{r})$ is now evaluated  by solving the GL equation. 
By assuming that the magnetic field points in the $z$ direction, 
$\mathbf{B}= B \mathbf{e}_z$, then the vector potential in the Landau gauge,  $\mathbf{A}=B x \, \mathbf{e}_y$, 
and the GL equation becomes:
\begin{equation}\label{GL1}    
\frac{-\hbar^2}{2m^*}  \left( \nabla_\perp - \frac{ie^* }{\hbar} Bx \mathbf{e}_y\right)^2  \Psi_0  -\left(a_T +b \vert \Psi\vert^2\right) \Psi_0 =0,
\end{equation}
where $m^*=2m_{ab}$ is the effective CP mass along the $ab$-plane. By taking into account that the  underlying crystal structure induces a square translational symmetry, Abrikosov showed \cite{abrikosov2004nobel} that Eq.(\ref{GL1}) admits an analytic solution determined by a Jacobi theta function:
\begin{equation}\label{abrikosov}
    \Psi_0(\tilde{x},\tilde{y})=C \exp[-\frac{1}{2}\kappa^2 \tilde{x}^2]
    \theta_3\left[1;\sqrt{2\pi} \ \kappa i (\tilde{x}+i\tilde{y}) \right],
\end{equation}
where  $\tilde{x}=x/\lambda_{ab}$, $\tilde{y}=y/\lambda_{ab}$, and the $ab$-penetration length, $\lambda_{ab}(T)\sim (n_s^{2D}(\mathbf{0},T)/m_{ab})^{-1/2}$. 
By writing $\Psi_0(\tilde{\mathbf{r}})=\vert \Psi_0(\tilde{\mathbf{r}})\vert e^{i\chi(\tilde{\mathbf{r}})}$ it follows that the GL current density is given by 
\begin{equation}\label{J}
\mathbf{J}_s=(\hbar e/m_{ab})\vert \Psi_0\vert^2(\nabla\chi-(2e/\hbar c)\mathbf{A}).
\end{equation}
Fig.(1) depicts the equiprobability contours lines defined by $\vert\Psi_0(\tilde{\mathbf{r}})\vert^2$, showing the emergence of a  periodic order parameter defining elementary  cells with period $\tilde{\mathcal{L}_x}=\tilde{\mathcal{L}_y}=\sqrt{2\pi}/\kappa$.  SC current lines  coincide with lines of constant $\Psi_0$, with the vortex cores located at the darkest zones. Notice that the vortex size can be tuned  by the substrate temperature $T_{sub}$ which modulates  the lattice parameter; in normal units to $\mathcal{L}_x(2 \, \mathrm{K}) \approx 6$ nm, whereas $\mathcal{L}_x(90 \, \mathrm{K}) \approx 100$ nm (see Supplementary Fig.~S1). 

We now assess the effect of the Abrikosov lattice on the NFRHT using Rytov's theory of fluctuating electrodynamics \cite{vinogradov2009thermally}. In this theory, the  spectral heat flux is determined by the ensemble average of the Poynting's vector: $\langle \mathbf{S}(\mathbf{r}) \rangle_\omega= 
(c/8\pi)\langle \mathbf{E}(\mathbf{r}) \times \mathbf{B}^*(\mathbf{r})\rangle_\omega + CC$. 
Denoting by $\lambda$ to the relevant electromagnetic wavelengths for the heat flux, for our system, $R/L<1$ and $R/\lambda\ll1$;  we accordingly assume that a dipolar interaction can safely describe the electromagnetic coupling between the sphere and the substrate \cite{tschikin2012,Manjavacas}. 

For a given separation, $L$,  the heat flux, $Q$, is then obtained from the integrated spectral distribution \cite{Vinogradov}, 
\begin{equation}
	Q(\mathbf{r},T_{sub},T_{ns},L)=\int_0^\infty d\omega \,  S_\omega(\mathbf{r},T_{sub},T_{ns},L),
\end{equation}
where
$S_\omega(\mathbf{r},T_{sub},T_{ns},L)=  \Delta \Theta_\omega( T_{ns},T_{sub})\, \rho_\omega(\mathbf{r},L) $, 
$\Delta \Theta_\omega( T_{ns},T_{sub})=\Theta_\omega(T_{ns})-\Theta_\omega(T_{sub})$, and the Planckian distribution, $\Theta_\omega(T) = \hbar \omega/[\exp(\hbar \omega/k_B T )-1]$. On the other hand, $\rho_\omega(\mathbf{r},L,T)=\rho^{prop}_\omega(\mathbf{r},L,T)+\rho^{evan}_\omega(\mathbf{r},L,T) $ represents the electromagnetic mode density contributions of propagating and evanescent waves to the heat flux, respectively. They are specified by $\varepsilon_{ns}$, $\varepsilon_{sub}$, and the nanosphere polarizability, $\alpha_{ns}=4\pi R^3 (\varepsilon_{ns}-1)/(\varepsilon_{ns}+2)$. Their explicit form is presented in SI.

\begin{figure}[t]
	\includegraphics[width=0.45\textwidth]{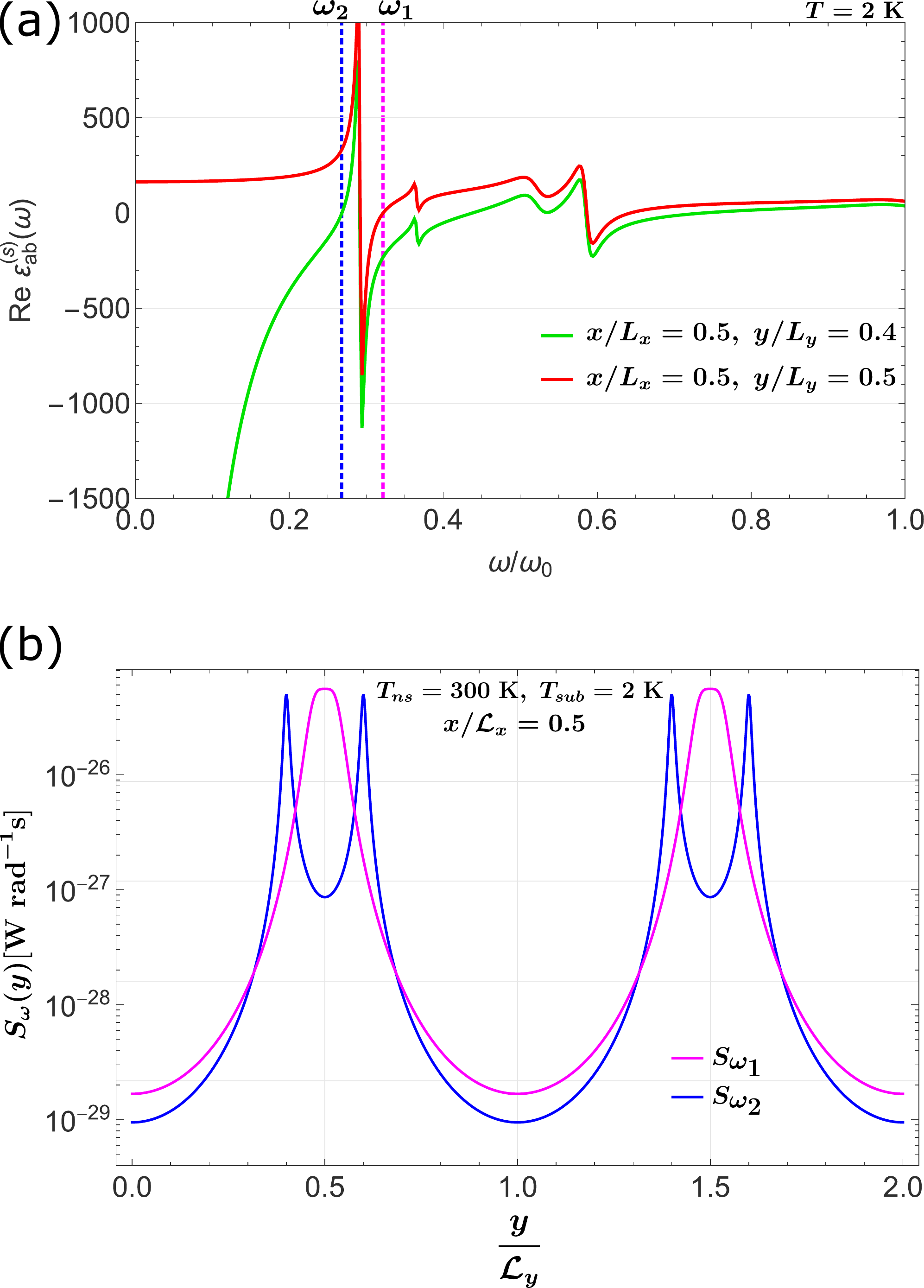}
	\caption{(a) Real part of the YBCO permittivity at $T=2$ K as a function of the normalized frequency $\omega/\omega_0$, $\omega_0=10^{14}$ rad/s.  Each curve corresponds to a different position in the Abrikosov lattice, marked by asterisks in Fig.~\ref{fig-02}~(b). 
	Vertical dashed lines indicate frequencies corresponding to the SM excitation condition, $\varepsilon_{ab}^{(s)}(\omega)=-1$, met for $\omega_1=0.322 \, \omega_0$ and $\omega_2= \, 0.268 \omega_0$, respectively.
	(b) Spatial dependence of the heat flux $S_\omega(y)$ for two different values of $\omega$ indicated as dashed vertical lines in Fig.~\ref{fig-03}~(a). Here $x/L_x=0.5$, $T_{ns}=300$ K and $T_{sub}=2$ K.}
	\label{fig-03}
\end{figure}

Figure~\ref{fig-02} shows the local radiative heat exchange between the YBCO nanosphere ($R=10$ nm, $T_{ns}=300$ K) and the SC YBCO substrate ($B\lesssim B_{c2}$) at two different temperature regimes, (a) $T_{sub}=90~\mbox{K}$, (b) $T_{sub}=2~\mbox{K}$, for $L=20~\mbox{nm}\ll \lambda_T$. The heat flux is normalized to the total heat flux $Q_{B=0}$ arising in the absence of the magnetic field when the substrate behaves as a homogeneous superconductor with  dielectric function $\varepsilon_{ab}^{(h)}$ ($i.e.$, $\vert \Psi_0\vert^2=1$ in (\ref{eps})). 
This figure reveals that the otherwise homogeneous heat flux acquires a periodic spatial dependence when the substrate supports the Abrikosov vortex lattice. It has been formerly shown that the SC transition causes a drop in the NFRHT as a consequence of the suppression of low-frequency SMs, which cannot be efficiently excited by thermal photons \cite{castillo2020near,castillo2022enhancing}. On the contrary, within the vortexes, the local density of SC carriers drops, and the presence of unscreened normal charge carriers causes an enhanced heat exchange.  

When $T_{sub}\lesssim T_c$, the density of SC carriers is low and the presence of the Abrikosov lattice is curtailed. As shown in Fig.~\ref{fig-02}~(a), $Q$ reaches its maximum at the center of the vortexes and exhibits a slight spatial modulation on the order of few percent. On the other hand, in the low-temperature regime, at $T_{sub}=2~\mbox{K}$, a 6-fold contrast increase between the minimum and maximum values of $Q$ across the unit cell can be observed. Notably, in this case the heat flux maximum suffers a radial displacement towards the vortex core borders forming a local basin, in contrast with what would be naively expected by solely considering the spatial distribution of $n_s^{2D}$. 
This puzzling behavior can be understood by considering that the relative weight of SMs that mainly contribute to the spectral heat flux can show a strong variation with the lattice location.
In principle, all modes associated to all frequencies participate to the heat transfer at each spatial point of the Abrikosov lattice unit cell. However, in order to qualitatively explain the observed spatial structure of the heat flux, we shall restrict the discussion to those	frequencies for which SMs are excited, expecting them to dominate the radiative properties of the lattice.
 Figure~(\ref{fig-03})(a)  portrays the real part of $\varepsilon_{ab}^{(s)}$ as a function of the normalized frequency $\omega/\omega_0$, with $\omega_0=10^{14}$ rad/s,  for two relevant points corresponding to the vortex center ($y=0.5\mathcal{L}_{y}$), and the vortex core border ($y \approx 0.4\mathcal{L}_{y}$) where, according to  Fig.~\ref{fig-02}, $Q$ is maximum. At these positions, the condition for SM excitation,  $\varepsilon_{ab}^{(s)}(\omega)=-1$, is met for $\omega_1=0.322 \, \omega_0$ and $\omega_2=0.268 \, \omega_0$, respectively. The corresponding spectral heat flux along the $y$-direction, $S_\omega(y)$,   is plotted in Figure~(\ref{fig-03})(b) for fixed $x=0.5\mathcal{L}_{x}$,  confirming that $S_{\omega 1}(y)$ attains its maximum value at the vortex core, while $S_{\omega 2}(y)$ reaches it at the core border.  
We verify in the SI that the contribution to the spectral heat flux from  other SM frequencies is less relevant due to the increased mode damping.
Finally, outside the vortexes $S_\omega$ is low because SMs cannot be excited at low frequencies due to the high reflectivity of the YBCO substrate.

The former results have been derived from a mean-field approach that neglects thermal fluctuations. When these are considered, vortex matter theories predict various states, including liquid, glassy, and nearly ordered crystalline structures with considerable spatial periodicity. At high temperatures and near $B_{c2}$ vortex matter may form a liquid state, whereas by lowering the temperature, it may freeze into a glassy state. Thus, the low-temperature regime discussed before seems to constitute the most favorable scenario to observe the spatial modulation of NFRHT.

In conclusion, we calculated the near-field radiative heat transfer properties of vortex material.
The quantum vortexes enhance the heat flux in their vicinity by breaking the superconducting state.
The main result of this letter is the prediction of the spatial structure of the heat flux arising from the vortex lattice on the superconducting surfaces.
Imaging spatial dependence of the near-field heat flux is possible using a room-temperature nanoprobe raster scanning a cold superconducting substrate subject to an external magnetic field.
Our results also suggest that each vortex acts as a heat transfer channel through which the energy exchange between the nanoprobe and the substrate in the near-field is selectively enhanced. 
We expect that by sweeping the magnetic field, it should be possible to track the evolution of the vortexes and the formation of the Abrikosov lattice.
Abrikosov vortexes have found several applications leading to the field of vortex matter \cite{blatter1997vortex}, where Majorana states near vortexes have been predicted \cite{akzyanov2017majorana}; likewise, they have been proposed as quantized information bits for use in digital cryoelectronics \cite{golod2015single}.
 We envision that the Abrikosov lattice can serve as a means to create ordered arrays of thermal sources at the nanoscale, opening a new direction of research in vortex materials. 
 
 \section{Acknowledgments}
G.P. acknowledges financial support from UNAM DGAPA PAPIIT Grant No. IN104522, UNAM DGAPA PAPIME PE101223 and CONACyT Projects No. 1564464 and No. 1098652. R. E-S acknowledges partial support from CONACyT Project No. A1-S-10537.
The authors acknowledge the valuable comments of R. Guti\'errez-J\'auregui. They also acknowledge graphical design by R. P\'erez Gonz\'alez.

\end{document}